\def\erg{{\rm\thinspace erg}}

\def\keV{{\rm\thinspace keV}}

\def\Msun{\hbox{$\rm\thinspace M_{\odot}$}}

\def\s{{\rm\thinspace s}}

\def\ergps{\hbox{$\erg\s^{-1}\,$}}

\documentclass[usegraphicx,usenatbib]{mn2e}
\usepackage{pdf14,times}
\usepackage{fixltx2e}
\usepackage{mathptmx}
\usepackage{multirow}
\usepackage{lscape}

\begin{document}

\title[Properties of AGN coronae II]{Properties of AGN coronae in the
  \textit{\textit{NuSTAR}} era II: hybrid plasma}
\author[A.C. Fabian et al] {\parbox[]{6.5in}{{
A.C.~Fabian$^1\thanks{E-mail: acf@ast.cam.ac.uk}$, A.~Lohfink$^1$,
R. Belmont$^{2,3}$, J. Malzac$^{2,3}$ and P. Coppi$^4$}\\ 
\footnotesize $^1$ Institute of Astronomy, Madingley Road, Cambridge
CB3 0HA\\ 
$^2$ Universite de Toulouse; UPS-OMP; IRAP; Toulouse,
France\\
$^3$ CNRS; IRAP; 9 Av. colonel Roche, BP 44346, F-31028 Toulouse cedex 4,
France\\
$^4$ Yale Center for Astronomy and Astrophysics, Yale University, New
Haven, Connecticut 06520-8121, USA  \\
 }}
\maketitle

%%%%%%%%%%%%%%%%%%%%%%%%%%%%%%%%%%%%%%%%%%%%%%%%%%%%%%%%%%%%%%%%%%%%%%%%
%%%%%%%%%%%%%%%%%%%%%%%%%%%%%%%%%%%%%%%%%%%%%%%%%%%%%%%%%%%%%%%%%%%%%%%%
%%%%%%%%%%%%%%%%%%ABSTRACT%%%%%%%%%%%%%%%%%%%%%%%%%%%%%%%%%%%%%%%%%%%%%%  
%%%%%%%%%%%%%%%%%%%%%%%%%%%%%%%%%%%%%%%%%%%%%%%%%%%%%%%%%%%%%%%%%%%%%%%%
%%%%%%%%%%%%%%%%%%%%%%%%%%%%%%%%%%%%%%%%%%%%%%%%%%%%%%%%%%%%%%%%%%%%%%%%
  
\begin{abstract}
The corona, a hot cloud of electrons close to the centre of the accretion
disc, produces the hard X-ray power-law continuum commonly seen in
luminous Active Galactic Nuclei (AGN).  The continuum has a
high-energy turnover, typically in the range of one to several 100 keV
and is suggestive of Comptonization by thermal electrons.  We are
studying hard X-ray spectra of AGN obtained with NuSTAR after
correction for X-ray reflection and under the assumption that coronae
are compact, being only a few gravitational radii in size as indicated
by reflection and reverberation modelling. Compact coronae raise the
possibility that the temperature is limited and indeed controlled by
electron-positron pair production, as explored earlier (Paper I).
Here we examine hybrid plasmas in which a mixture of thermal and
nonthermal particles is present.  Pair production from the nonthermal
component reduces the temperature leading to a wider 
temperature range more consistent with observations.
\end{abstract}


%%%%%%%%%%%%%%%%%%%%%%%%%%%%%%%%%%%%%%%%%%%%%%%%%%%%%%%%%%%%%%%%%%%%%%%%
%%%%%%%%%%%%%%%%%%%%%%%%%%%%%%%%%%%%%%%%%%%%%%%%%%%%%%%%%%%%%%%%%%%%%%%%
%%%%%%%%%%%%%%%%%%INTRO%%%%%%%%%%%%%%%%%%%%%%%%%%%%%%%%%%%%%%%%%%%%%%  
%%%%%%%%%%%%%%%%%%%%%%%%%%%%%%%%%%%%%%%%%%%%%%%%%%%%%%%%%%%%%%%%%%%%%%%%
%%%%%%%%%%%%%%%%%%%%%%%%%%%%%%%%%%%%%%%%%%%%%%%%%%%%%%%%%%%%%%%%%%%%%%%%

\section{Introduction}

The primary power-law, X-ray emission from luminous accreting black
holes originates from the corona, which is a hot cloud close to the
centre of the
accretion disc.  Soft disc photons are Compton-upscattered in the
corona to make the power-law continuum. It often shows a high-energy
turnover somewhere between about 30 keV and 1 MeV, generally interpreted
as the temperature of the corona (or a value close to it). The primary
continuum is usually the most variable part of the spectrum, showing
rapid changes which, when interpreted using X-ray reflection and
reverberation signatures, indicate that the corona has a relatively small
size of less than a few tens of gravitational radii ($r_{\rm
  g}=GM/c^2$) \citep{Fabian2013,Reis2013,Uttley2014}. The fraction of
the bolometric power passing through the corona ranges from about 10
to 50 per cent \citep{Vasudevan2007}.

A compact corona in a luminous accreting black hole system must be a
dynamic structure since the heating and cooling timescales for the hot
electrons are less than the light crossing time of the corona
\citep[Paper I, F15]{Fabian2015a}. There is not enough time for any
simple equilibrium to be established. This can be further considered
by looking at the thermal energy content of the corona, which is small
\citep[MF01]{Merloni2001}.  A spherical corona with typical properties
of unit Thomson depth, $\tau_{\rm T}=1$,  temperature
$kT=50\keV$ and radius $10R_1 r_{\rm g}$ above a $10^6m_6\Msun$ black
hole has a very small mass of $M_{\rm c}=10^{-8}m_6^2R_1^2\Msun$,
thermal energy
\begin{equation}
E_{\rm th}\approx 10^{42} m_6^2R_1^2 \erg
\end{equation}
and crossing time $t_{\rm cross}=50m_6R_1\s$. The luminosity obtained
if all the coronal energy is released on a crossing time
$L_{\rm cross}\approx 2\times 10^{40}m_6R_1\ergps$is much less than
the Eddington luminosity $L_{\rm Edd}$, in that
$L_{\rm cross}/L_{\rm Edd}\approx10^{-4}R_1$. Higher luminosities, as
are typically observed, require that both cooling and heating occur on
a faster timescale.  As argued by MF01, the energy supplying the heat
must be stored in the corona, probably in the form of magnetic fields
powered by the accretion disc. It is then unclear, if only heating and
cooling are involved, what controls the plasma temperature, or
temperatures, in the corona and how any stable equilibrium is
established.

Here we follow up on our recent paper (F15), which explored the properties
(temperature and compactness) of coronae  deduced in
the light of \textit{NuSTAR} observations. The coronal plasma is
assumed to be thermal and the temperature at the maximum allowed by
electron-positron pair production: heating pushes the temperature
upward until it is balanced by the creation of pairs -- the pair
thermostat \citep{Svensson1984,Zdziarski1985}. \textit{NuSTAR}
observations of Active Galactic Nuclei (AGN) show they avoid the
region where pair production runs away, indicating that the process
may play a role. However there is a significant number of AGN for
which the temperature is much smaller than expected on the basis of
purely thermal pair production \citep[e.g.,][]{Balokovic2015,Ursini2016}.

Here we aim to tackle the issue of a low temperature corona by
considering hybrid coronae containing a mixture of
thermal and nonthermal particles. We assume that the corona is highly
magnetized and powered by dissipation of magnetic energy. Such
dissipation is often intermittent in both space and time.  In a
compact corona the heating and cooling are so intense that energetic
particles may not have time to thermalize before inverse Compton
cooling reduces their energy (see the electron-electron equilibration
line in Fig.~1, which is adapted from F15: the electron-proton line lies
about 3 orders of magnitude lower). Only a small fraction of
the electrons need start from energies of an MeV or more for them to
emit hard photons which collide and create electron-positron pairs. If
the pairs are energetic enough then yet more pairs are produced and a
runaway situation can occur. Meanwhile the cooled pairs can, before
annihilating, share the available energy leading to a reduction in the
mean energy per particle and thus  temperature of the thermal
population, which may be composed mostly of pairs. This can result in
a relatively cool thermal particle population with a temperature well
below 100 keV, a low-level hard tail and a possible broad annihilation
line at 511 keV. Unless very sensitive hard X-ray observations are
obtained, the emitted continuum appears to originate from a
low-temperature thermal plasma incapable of pair
production. Paradoxically, what appear to be the lowest temperature
objects can be the most pair-dominated.

Hybrid models for the corona were introduced over two decades ago
\citep{Zdziarski1993,Ghisellini1993} providing a better explanation
for the hard X-ray spectra of luminous accreting black holes than
purely nonthermal \citep{Svensson1994} or thermal \citep{Fabian1994}
models. Comparison of hybrid models with AGN spectra has been carried
out on NGC4151 \citep{Zdziarski1996,Johnson1997}, using OSSE data from
the Compton Gamma-Ray Observatory to conclude that the nonthermal
contribution in NGC4151 is less than 15 per cent. A study of stacked
AGN spectra from OSSE by \citet{Gondek1996} was unable to distinguish
between the possibilities. More recent work on NGC4151 using INTEGRAL
data \citet{Lubinski2010} shows that thermal models fit the spectra,
which extend up to 200 keV, in its bright state and that the source
appears harder in its low state. There is no new information on any
possible hard tail or annhilation feature; OSSE results remain the
most sensitive at those energies.  Integral spectra in the 2--200 keV
band of 28 bright Seyfert galaxies, combined with lower energy data
from other instruments, are characterized by thermal Componization
with a mean temperature $kT\sim50$\,keV \citep{Lubinski2016} and a
tail to higher values. We found a similar distribution from NuSTAR
data (F15, Fig.~1).

\begin{figure} \centering
  \includegraphics[width=0.99\columnwidth,angle=0]{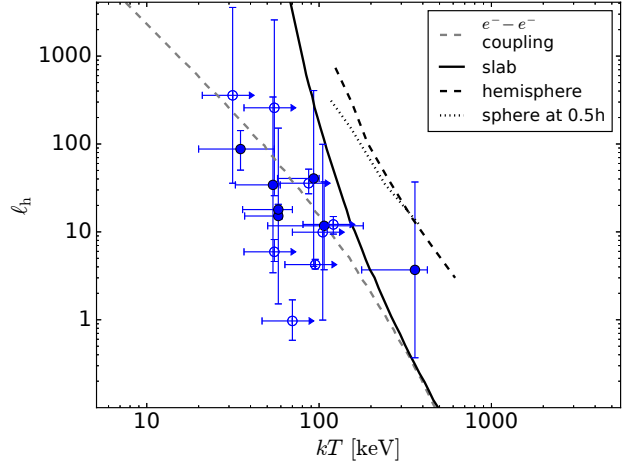}
  \caption{Electron temperature ($kT$) -- compactness ($\ell$)
    distribution for \textit{NuSTAR} observed AGN (blue points) as
    first shown in \citet{Fabian2015a}. The e--e coupling line from
    \citet{Ghisellini1993} is included: electrons in the region above
    the dashed line do not have time to thermally equilibrate with
    each other before cooling occurs. Pair lines, above and to the
    right of which there is runaway pair production, taken from
    \citet{Stern1995} are also shown. The slab line has been
    extrapolated slightly to higher $\ell$. Open circles are used for
    objects where the temperature measurement is a lower limit. The
    temperatures shown are the observed values reported from spectral
    fits.  }
\end{figure}

Hybrid models have been successfully applied to the spectra of X-ray
Black Hole Binaries (BHB) such as Cyg X-1 where high energy tails are
clearly seen \citep{Gierlinski1999,McConnell2001,Parker2015}, for
other sources see also \citep{Grove1998,Kalemci2016, Wardzinski2002,Droulans2010}. In this paper, we concentrate on AGN, since
high-frequency, X-ray reverberation results indicate the location and
size of the corona. Such reverberation has not yet been detected in
BHB.

Our goal here is to compare the range of NuSTAR results with the pair 
thermostat operating in hybrid plasma. We are not at this stage
attempting to reproduce spectra of individual objects. We 
define our parameters and discuss hybrid plasmas in more detail in
Section 2. We compute and show $kT-\ell$ diagrams for various
levels of the nonthermal fractional contribution. Finally we discuss
our conclusions in Section 3. Detailed spectral comparison with
individual objects is left to a later paper.

\section{Hybrid plasmas}
The excellent hard X-ray sensitivity of \textit{NuSTAR} has allowed
detailed spectral modelling of the reflection component and precise
measurements of the spectral turnover in the continuum spectrum of
many X-ray bright Active Galactic Nuclei (AGN) to be made.

\begin{figure}
  \includegraphics[width=0.99\columnwidth,angle=0]{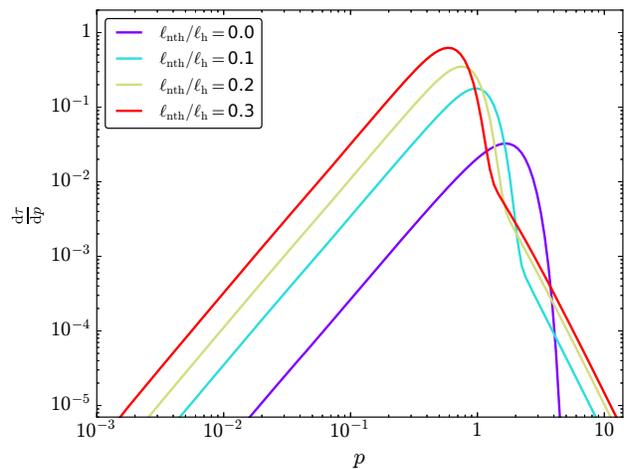}
  \caption{Lepton distributions for an increasing sequence of
$\ell_\mathrm{nth}/\ell_\mathrm{h}$ and $\ell_h/\ell_s=1$ at
$\ell_{\rm h}=100$, where $p$
is the dimensionless lepton momentum ($\beta\,\gamma$) and
$\frac{\mathrm{d}\tau}{\mathrm{d}p}=R\times\sigma_\mathrm{T}\times\frac{\mathrm{d}n}{\mathrm{d}p}$
with $R$ being the radius of the sphere modelled by Belm,
$\sigma_\mathrm{T}$ the Thomson cross-section, and $n$ the lepton
number density. As the non-thermal fraction increases the peak of the
distribution moves to lower energies.}\label{lepspec}
\end{figure}

\begin{figure*}
  \begin{minipage}{0.49\textwidth}
    \includegraphics[width=0.99\columnwidth,angle=0]{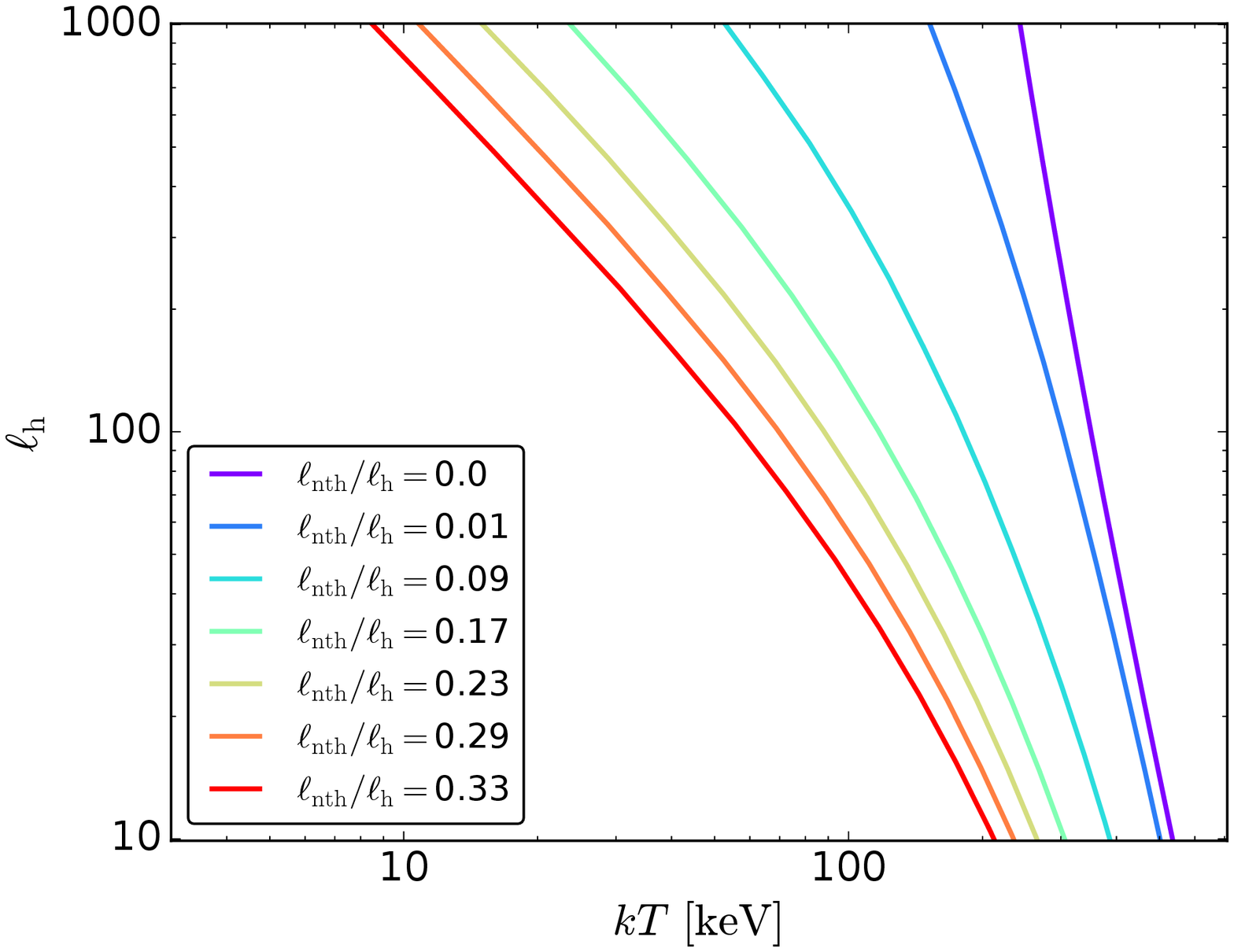}
  \end{minipage} \hfill
  \begin{minipage}{0.49\textwidth}
  \includegraphics[width=0.99\columnwidth,angle=0]{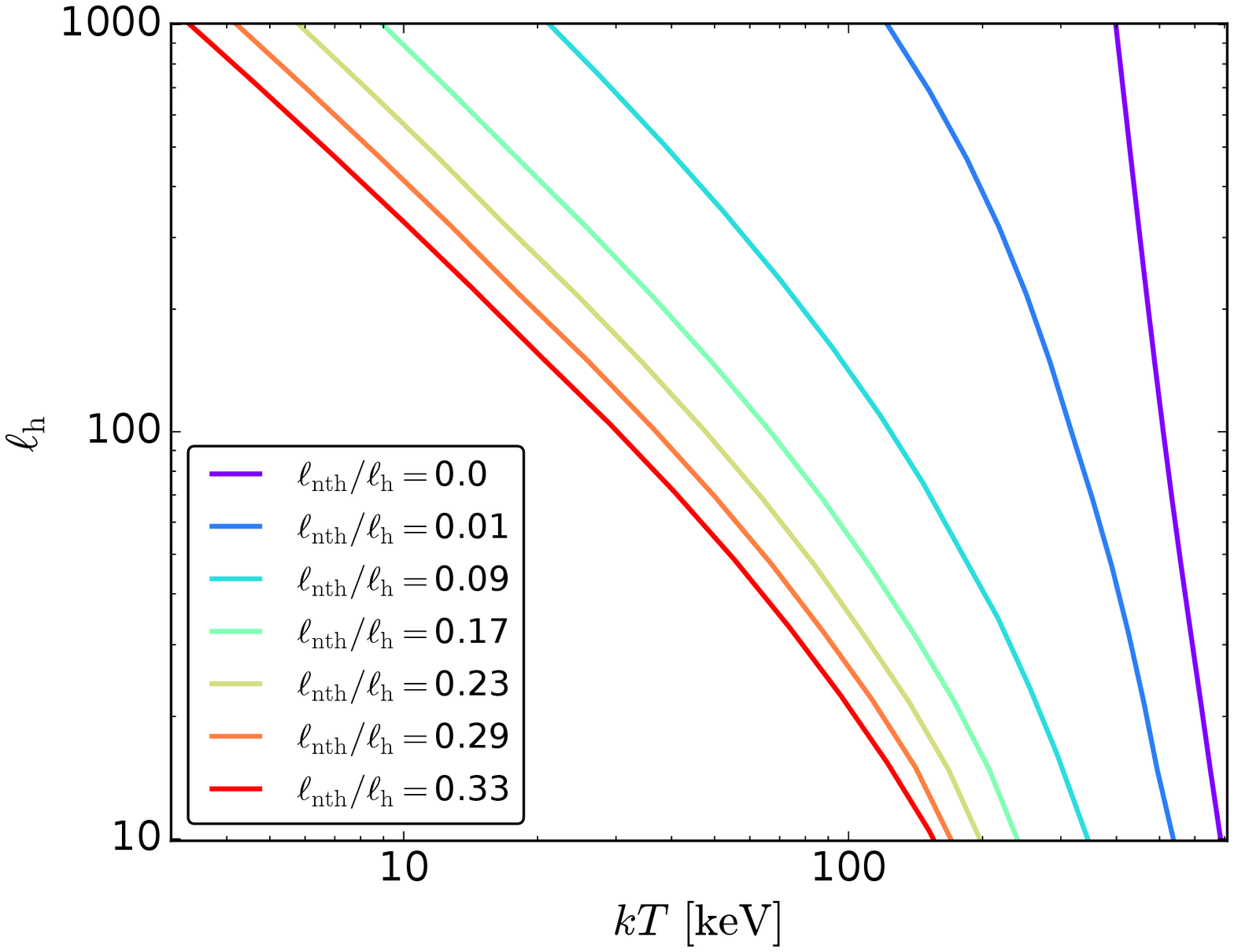}
  \end{minipage}
  \caption{$kT-\ell$ distributions for an increasing sequence of
$\ell_\mathrm{nth}/\ell_\mathrm{h}$ and two different $\ell_h/\ell_s$
values: 1 (left panel), 0.1 (right panel). As the non-thermal fraction
increases the equilibrium temperature is lower. Similarly, a geometry
with more soft photons entering the corona leads to lower equilibrium
temperatures.}
\end{figure*}

In order to understand what sets those temperatures, it is useful to
show these measurements in the form of a temperature-compactness,
$kT-\ell$, diagram (Fig.~1; adapted from F15), where $kT$ is the
electron temperature and $\ell$ is the dimensionless compactness
parameter
\begin{equation} \ell=\frac{L}{R}\frac{\sigma_{\rm T}}{m_{\rm
e}c^3}\qquad
\end{equation} with $L$ being the luminosity, $R$ the radius of the
source (assumed spherical), $\sigma_{\rm T}$ the Thomson cross section
and $m_{\rm e}$ the mass of the electron.

Coronae should lie along one of the lines in the $kT-\ell$ plot, which
in Fig.~1 represent the expectation for a purely thermal plasma, whose
temperature is regulated by pairs. This takes place as the corona is
heated and driven diagonally upwards from left to right in the plot
(i.e. increasing both $L$ and $kT$) until pairs start to form from
energetic collisions. This then reduces and finally stabilizes the
temperature at the line.

Sources should not be seen to the right of the line, since
catastrophic pair production would ensue, rapidly driving down the
temperature. If a source is confirmed to occur and remain in the
forbidden region, then the basis of the model needs to be
re-considered, with something other than a static compact corona being
relevant. Jetted emission would be more likely in this case. Objects
well to the left of the line pose the problem of what stabilizes them
in such a highly dynamic situation?

\begin{figure}
  \includegraphics[width=0.99\columnwidth,angle=0]{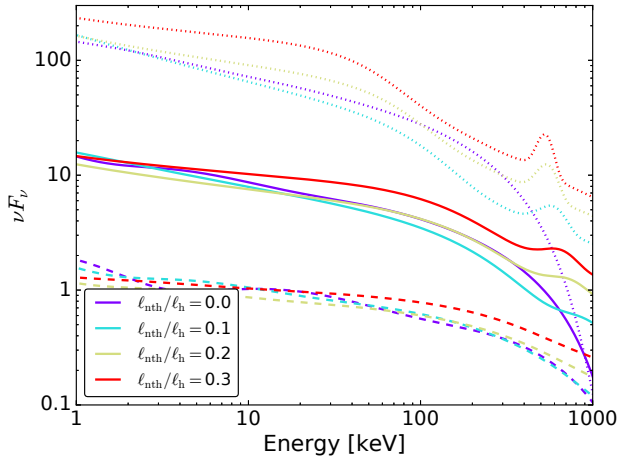}
  \caption{Simulated spectra for $\ell_{\rm h}$ of 10, 100, 1000
(dashed, solid, dotted) and $\ell_{\rm h}/\ell_{\rm s}$ of 1 for an
increasing sequence of $\ell_{\rm nth}/\ell_{\rm h}$. }
\end{figure}

\begin{figure}
  \includegraphics[width=0.99\columnwidth,angle=0]{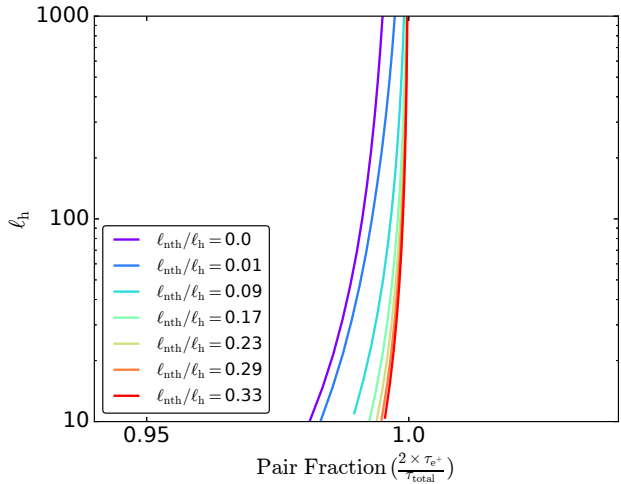}
  \caption{Pair fraction for an increasing sequence of
$\ell_\mathrm{nth}/\ell_\mathrm{h}$ and $\ell_h/\ell_s=1$. While the
corona can be considered pair dominated for all cases, the pair
fraction increases further for a higher non-thermal contribution. }
\end{figure}

We propose here that hybrid coronae can solve this problem. In our
hybrid corona scenario, we assume that part of the coronal heating
power, $L_{\rm nth},$ creates a nonthermal power-law electron spectrum
within the source, with compactness parameter $\ell_{\rm nth}$. The
thermal heating of the electron population $\ell_{\rm th}$ and the
injection of soft photons $\ell_{\rm s}$ continues as in a standard,
thermal scenario. The total heating is then given by
$\ell_{\rm h}=\ell_{\rm th}+\ell_{\rm nth}$ and the nonthermal
fraction is thus $\ell_{\rm nth}/\ell_{\rm h}$.  The ratio
$\ell_{\rm h}/\ell_{\rm s}$, together with the Thomson optical depth
$\tau$, plays a strong role in determining the power-law slope of the
emerging X-ray radiation, with high values leading to a flat spectrum
and low values to a steep one. Increasing $\tau$, which can be a
result of pair production, tends to harden the spectrum.

We use the {\sc BELM} code \citep{Belmont2008} for simulating emission
from hybrid plasmas. (Comparable results are obtained using {\sc
EQPAIR}\footnote{An excellent discussion of hybrid plasmas and their
possible application to BHB is given in the unpublished paper by Paolo
Coppi at http://www.astro.yale.edu/coppi/eqpair/eqpap4.ps}
\citep{Coppi1999}.)  {\sc BELM} simultaneously solves the coupled
kinetic equations for leptons and photons in a (magnetized,) uniform,
isotropic medium. The equilibrium energy distributions of the
Comptonizing leptons are calculated self-consistently, taking into
account the coupling between particles and radiation. Radiation
transfer is dealt with using a simple escape probability formalism.
The code takes into account all the relevant radiation processes such
as self-absorbed synchrotron radiation, Compton scattering,
self-absorbed bremsstrahlung radiation, pair production/annihilation,
coulomb collisions, and uses prescriptions for particle
heating/acceleration.  In the simulations shown in this paper the
effects of cyclo-synchrotron radiation are neglected (magnetic field
is set to 0 G).  {\sc EQPAIR} is very similar (and gives identical
results) but assumes the low energy component of the electron
distribution is a perfect Maxwellian, while our electron distributions
are calculated according to the full kinetic equations in {\sc BELM}.
Non-thermal particles are injected with a power-law index of 2.5
ranging from $\gamma_1=1.3$ to $\gamma_2=1000$ for the purpose of this
paper.

The electron temperature is obtained by fitting the thermal part of
the lepton distribution with a Maxwellian. While this approximation
becomes less accurate as the non-thermal contribution to the lepton
distribution increases (Fig.~\ref{lepspec}), it still characterizes
the general behavior of the temperature well.

Figs~3 show the $kT-\ell$ distribution as lines at constant values of
$\ell_{\rm nth}/\ell_{\rm h}$ and for two different values (1 and 0.1)
of $\ell_{\rm h}/\ell_{\rm s}$.  Hard X-ray spectra for a fixed ratio
of $\ell_{\rm th}=100$ and an increasing sequence of $\ell_{\rm
nth}/\ell_{\rm h}$ are shown in Fig.~4.  We see that as the nonthermal
power is increased from zero, then the temperature of the thermal
component drops from around several hundred keV to below 20\,keV. The
drop in spectral turnover is apparent in the spectra as is the rise of
the hard X-ray flux and the annihilation line at the highest
nonthermal fractions.

To assure ourselves that the observed decrease in electron temperature
is due to pairs we plot the pair fraction in Fig~5. It is clear that
the coronae are pair dominated.

Finally, in Fig.~6 we show the \textit{NuSTAR} data points from Fig.~1
again in the $kT-\ell$ diagram with our hybrid corona results
overlaid. A few new data points have been added from very recent
results, see Table~1 for details. The new values were obtained in a
similar fashion to \citet{Fabian2015a}. In contrast to Fig.~1, we
corrected the datapoints for gravitational redshifting assuming the
measured coronal size and height given in Table~1 or $10r_{\rm g}$ if
no measurement exists. A 10 to 30 per cent nonthermal contribution
appears sufficient to account for most of these objects and
particularly those with well constrained temperatures.

We have not accounted for the effect of light bending on coronal flux
as this depends on the inclination and other parameters. It could
increase the intrinsic value of $\ell_h$ by a factor of two, or even
more if the corona lies close to the black hole. If the emission site
within the corona is very localized then the observationally inferred
value of $\ell_h$ could be underestimated by a factor of a few.

We note that most of the spectral fits used in determining the coronal
temperatures assume a power-law continuum with an exponential
high-energy cutoff as an approximation to a full Comptonization
spectrum. This introduces a systematic uncertainty in the temperature,
since the cutoff energy is 2--3 times higher than the gas temperature
depending on the optical depth of the corona (i.e whether thin or
thick) and also on the geometry of the corona and its illumination
\citep[see e.g.][]{Petrucci2001}.

\begin{figure} \centering
  \includegraphics[width=0.99\columnwidth,angle=0]{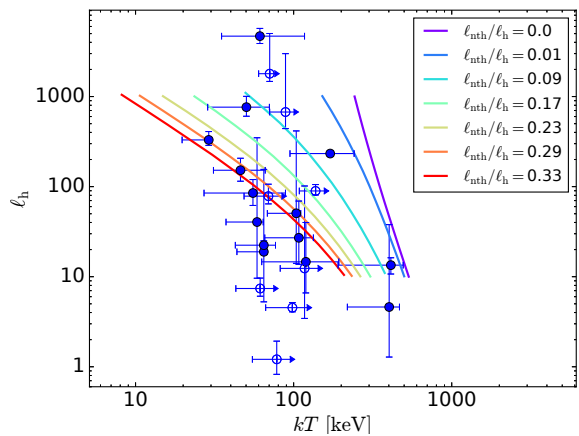}
  \caption{\textit{NuSTAR} data overplotted on the theoretical
predictions for $kT-\ell$ distributions for an increasing sequence of
$\ell_\mathrm{nth}/\ell_\mathrm{h}$ with $\ell_{\rm h}/\ell_{\rm
s}=1$. The data have been corrected for gravitational redshift.}
\end{figure}

\section{Discussion}

We find that current data are consistent with the pair thermostat
operating and determining the temperature of compact AGN coronae,
provided that a hybrid plasma is involved. For most sources, only a
small addition of nonthermal plasma is required.  The wide range of
temperatures at a fixed value of $\ell_h$ can be accounted for with a
non-thermal fraction ranging up to 30 per cent. Objects measured to
have the lowest temperatures require the greatest nonthermal
fraction. They should then have the strongest high energy tail and
broad annihilation lines. Such features are generally undetectable
with present instrumentation in AGN.
 
We also acknowledge that the simple one zone approach in the current
modelling is probably too simplistic to capture the likely geometry of
the corona and introduces a small uncertainty in the predicted
temperatures.

In future work, we shall explore the location of individual sources in
$kT - \ell_{\rm h} - \ell_{\rm h}/\ell_{\rm s}$ space, which plays a
role in determining the spectral index as well as the continuum
cutoff.  We shall also include Stellar Mass Black Hole Binaries (BHB)
and investigate the effect of magnetic fields acting through the
synchrotron boiler (see e,g. \citet{Malzac2008,Veledina2010}) on
current spectral results.

\section*{Acknowledgements} We thank the referee for helpful comments.
ACF and AL acknowledge support from ERC Advanced Grant FEEDBACK
340442. R. Belmont and J. Malzac acknowledge financial support from
the French National Research Agency (CHAOS project ANR-12-BS05-0009)

\bibliographystyle{mn2e} \bibliography{coronae2}

%\begin{thebibliography}{109} %\expandafter\ifx\csname
%natexlab\endcsname\relax\def\natexlab#1{#1}\fi
%\end{thebibliography}

\begin{table*}
\caption{The targets and properties of the active galactic nuclei with
cut-off constraints resulting from observations with
\textit{NuSTAR}. The references to the individual works are given in
the right-most column.}\label{data_nustar}
\begin{center}
\begin{tabular}{|c|c|c|c|c|c|c|c|c|c|c|c|c|} \hline Source & z &
$\log(M)$ & $r_{\rm co}$ & $F_x$ & $E_{\rm cut}$ & $kT$ & $\ell$ &
$(1+z_\mathrm{grav})$ & $\Gamma$ & Data & References \\ & &
[$M_\odot$] & [$r_G$] & & [keV] & & & & & \\ \hline \hline Cen\,A &
0.002 & $7.65_{-0.14}^{+0.11}$ & 10 & 34.34 & $>1000$ & $>333$ &
$11_{-3}^{+4}$ & 1.12 & $1.80\pm0.01$ & XMM/NU & $1-2$ \\ NGC\,4593 &
0.009 & 5.45$_{-0.15}^{+0.13}$ & 10 & 0.50 & 90$_{-20}^{+40}$ &
$45_{-22}^{+20}$ & $612_{-158}^{+242}$ & 1.12 & $1.59_{-0.02}^{+0.03}$
& XMM/NU & $3-4$ \\ Mrk\,766 & 0.013 & $6.8_{-0.06}^{+0.05}$ & 3.4 &
0.88 & $>441$ & $>147$ & $244_{-27}^{+34}$ & 1.48 &
$2.22_{-0.03}^{+0.02}$ & XMM/NU & $5-6$ \\ 3C\,120 & 0.033 &
$7.75\pm0.04$ & 10 & 2.23 & $305_{-74}^{+142}$ & $153_{-76}^{+71}$ &
$153_{-16}^{+18}$ & 1.12 & $1.70_{-0.03}^{+0.10}$ & XMM/NU & $7-8$ \\
PG\,1211+143 & 0.081 & 8.16$_{-0.16}^{+0.11}$ & 10 & 0.32 & $>124$ &
$>41$ & $63_{-14}^{+28}$ & 1.12 & $2.51\pm0.2$ & NU & $9-10$ \\
4C\,74.26 & 0.104 & $9.6\pm0.5$ & 7 & 1.17 & $183_{-35}^{+51}$ &
$92_{-42}^{+26}$ & $20_{-13}^{+42}$ & 1.18 & $1.81\pm0.03$ & Swift/NU
& $11-12$ \\ 3C\,273 & 0.158 & $8.84_{-0.11}^{+0.16}$ & 10 & 1.71 &
52$_{-2}^{+2}$ & $26_{-9}^{+1}$ & $365_{-43}^{+79}$ & 1.12 & $1.66\pm
0.01$ & NU & $13-14$ \\ PG\,1247+267 & 2.038 & 8.3$_{-0.15}^{+0.17}$ &
4 & 0.02 & 89$_{-34}^{+134}$ & $45_{-26}^{+56}$ &
$2480_{-803}^{+1023}$ & 1.38 & $2.35_{-0.08}^{+0.09}$ & XMM/NU &
$15-16$ \\ %GRS\,1734-292 & 0.021 & 89$\pm0.5$ & 10 & 2.36 &
53$_{-2}^{+2}$ & XX & XX & XMM/NU & $17-18$ \\ \hline
 \end{tabular}
\end{center}
\begin{flushleft} $F_x$ is the 0.1-200\,keV X-ray flux in
$10^{-10}\,{\rm erg}\,{\rm cm}^{-2}\,{\rm s}^{-1}$.\\
\textbf{References:} 1: \citet{Fuerst2015a}, 2: \citet{Neumayer2007},
3: \citet{Ursini2016}, 4: \citet{Peterson2005}, 5: Buisson et~al., in
prep., 6: \citet{Bentz2010},\citet{Grier2013}, 7: Lohfink et~al., in
prep., 8: \citet{Grier2012a}, 9: \citet{Zoghbi2015}, 10:
\citet{Peterson2004}, 11: Lohfink et~al., submitted, 12:
\citet{Woo2002}, 13: \citet{Madsen2015}, 14: \citet{Peterson2004}, 15:
\citet{Lanzuisi2016a} ,16: \citet{Trevese2014}%, 17:
\end{flushleft}
\end{table*}

\end{document}